\documentclass[aps,pre,twocolumn,groupedaddress,showpacs]{revtex4-1}
\usepackage{bm}
\usepackage{url}
\usepackage{amssymb}
\usepackage{pifont}
\usepackage{amsmath}
\usepackage{subfigure}

\usepackage{graphics}
\usepackage{graphicx}
\usepackage{epsfig}
\usepackage{pstricks}
\usepackage{pst-plot}
\usepackage{pst-slpe}
\usepackage{latexsym}

 \newcommand{\be}{\begin{equation}}
 \newcommand{\ee}{\end{equation}}

\begin{document}
\title{Percolation in a distorted square lattice}
\author{Sayantan Mitra}
\author{Dipa Saha}
\author{Ankur Sensharma}
\email{itsankur@gmail.com}

\affiliation{Department of Physics, University of Gour Banga, Malda - 732103, West Bengal, India}

\date{\today}


\begin{abstract}
This paper presents a Monte-Carlo study of percolation in a distorted square lattice, in which, the adjacent sites are
not equidistant. Starting with an undistorted lattice, the position of the lattice sites are shifted
through a tunable parameter $\alpha$ to create a distorted empty lattice. In this model, two neighboring sites
are considered to be connected to each other in order to belong to the same cluster, if both of
them are occupied as per the criterion of usual percolation and the distance between them is less
than or equal to a certain value, called connection threshold $d$.
While spanning becomes difficult in distorted lattices as is manifested by the increment of the percolation
threshold $p_c$ with $\alpha$, an increased connection threshold $d$ makes it easier for the system to
percolate. The scaling behavior of the order parameter through relevant critical exponents and the fractal
dimension $d_f$ of the percolating cluster at $p_c$ indicate that this new type of percolation may belong to the same
universality class as ordinary percolation. This model can be very useful in various realistic applications
since it is almost impossible to find a natural system that is perfectly ordered.

\end{abstract}
\maketitle
\section{Introduction}\label{Intro}
Percolation is an intensely studied model of statistical mechanics and has been widely applied to
interpret and describe numerous physical, natural and social systems \cite{book}. The popularity of the
model can be attributed to the coexistence of simplicity in its proposition and richness in its outcome \cite{review}.

The mathematical model of percolation was first proposed by Broadbent and Hammersley in 1957 \cite{Broadbent}. It gradually
became widely accepted by physicists \cite{Isichenko} and was successfully applied to study and understand the properties of
metal-insulator transition \cite{Ball}, magnetic materials \cite{Dotsenko}, spin quantum Hall effect \cite{Gruzberg},
growth models \cite{Saberi, Dsouza} and networks \cite{Derenyi, Callaway, Kalisky}.
Percolation is also frequently used in subjects like chemistry, geophysics, environmental sciences, medical
sciences and social sciences to analyze issues such as polymer gelation \cite{Coniglio}, colloids \cite{Anekal, Gnan},
flow of oil through
porous media \cite{King}, fractality of coast lines \cite{Sapoval, Saberi_coast}, spreading of forest fire
\cite{Albano, Albano2}, epidemic outbursts \cite{Grassberger}, neuron communication \cite{Zhou}, tumor induced
angiogenesis \cite{Paul} and numerous other systems. It is a highly active field of research with many open
problems \cite{Araujo}.

In a typical site percolation problem, the sites of a regular empty lattice are occupied randomly with a probability $p$,
called occupation probability. A cluster is formed when two neighboring sites are occupied. If any nearest
neighbor of any of the sites in the cluster gets occupied, it is also included in the cluster. For small
values of $p$, many small isolated clusters are formed in the lattice. As $p$ is gradually increased, these
clusters start to merge and at a certain value of $p$ ($=p_c$, called the percolation threshold), a single cluster
spans the lattice. This sudden appearance of a spanning cluster marks a phase transition (continuous in this case)
when the cluster-size and the correlation length diverge. The percolation transition possesses a number of
remarkable characteristic features and exhibits interesting critical behavior to form an important
universality class.

Another primitive and widely used model is bond percolation where the empty bonds between the preoccupied sites 
are occupied randomly.
The value of percolation threshold for bond percolation in infinite square lattice has been analytically
calculated to be $p_c=\frac{1}{2}$, unlike site percolation, for which, best numerical estimate is $p_c\approx0.592746$.
An interesting extension of these two basic models is the site-bond percolation \cite{Hoshen}, where the sites are occupied with
probability $p_s$ and the bonds between neighboring occupied sites are filled with probability $p_b$.
Percolation models like explosive percolation \cite{Achlioptas, Ziff_explo, Cho, Dsouza}, bootstrap percolation \cite{Adler},
directed percolation \cite{Broadbent, Grassberger, Albano},
correlated percolation \cite{Coniglio, Makse} and a lot of other variants are available in the literature. 
These models have been proposed and
studied not only for meeting the requirement of different systems but also out of pure mathematical interest. 

Several percolation studies have addressed the geometric and transport properties of disordered systems
\cite{Coniglio, Makse, Araujo2}. A model \cite{Kundu}
has been introduced with an additional source of disorder, in which the sites are occupied randomly with discs of
random radii. The bonds are considered occupied if the discs satisfy certain predefined conditions. The critical
behavior indicates that this model belongs to the same universality class as ordinary percolation. Another 
interesting model \cite{Hassan} deals with a weighted planar stochastic lattice (WPSL); and from the calculated
values of the critical
exponents the authors conclude that percolation on WPSL belongs to a different universality class.

 \begin{figure*}[t]
 \centering
 \subfigure[]{\includegraphics[scale=0.4]{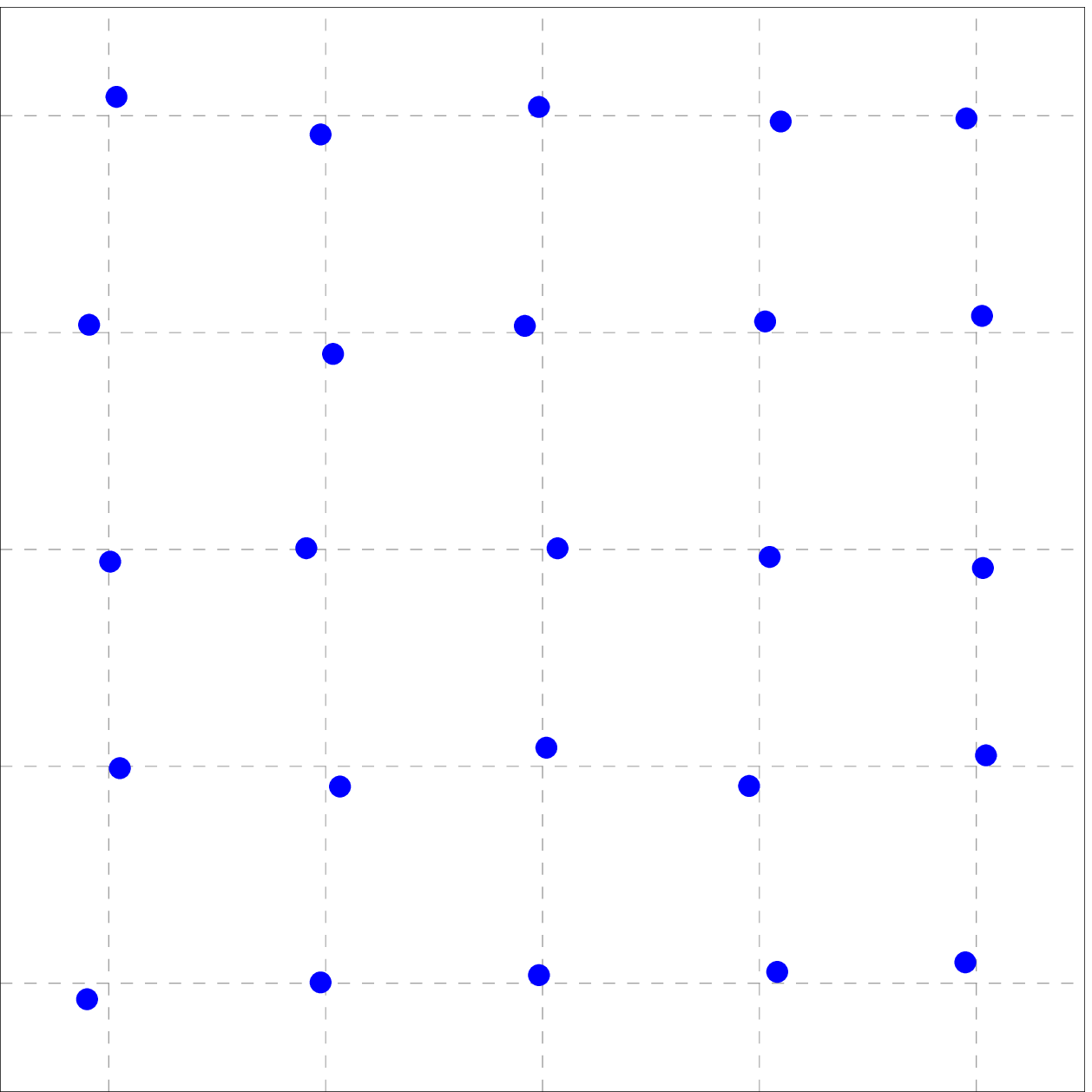}}\hspace{0.2cm}%
 \subfigure[]{\includegraphics[scale=0.4]{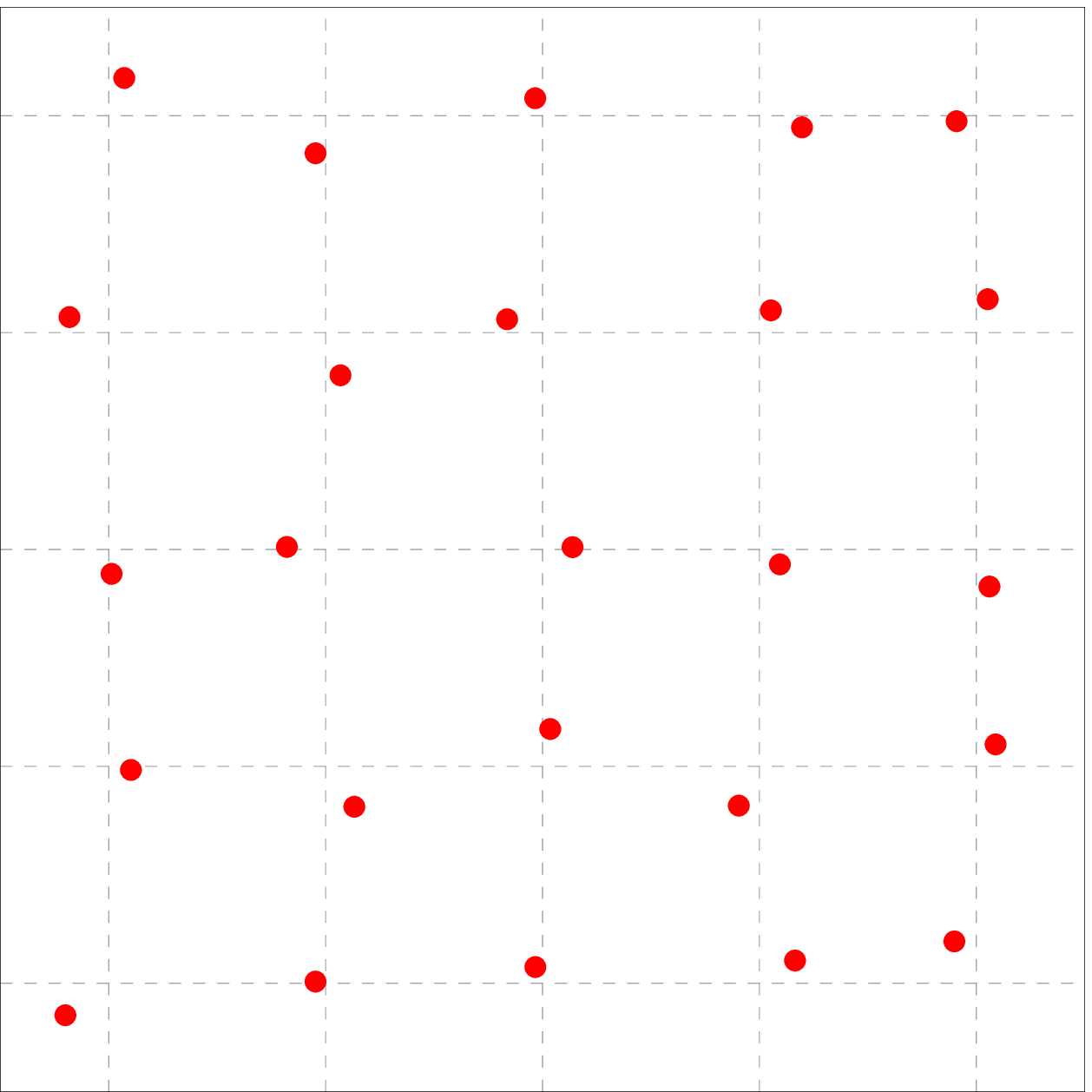}}\hspace{0.2cm}%
 \subfigure[]{\includegraphics[scale=0.4]{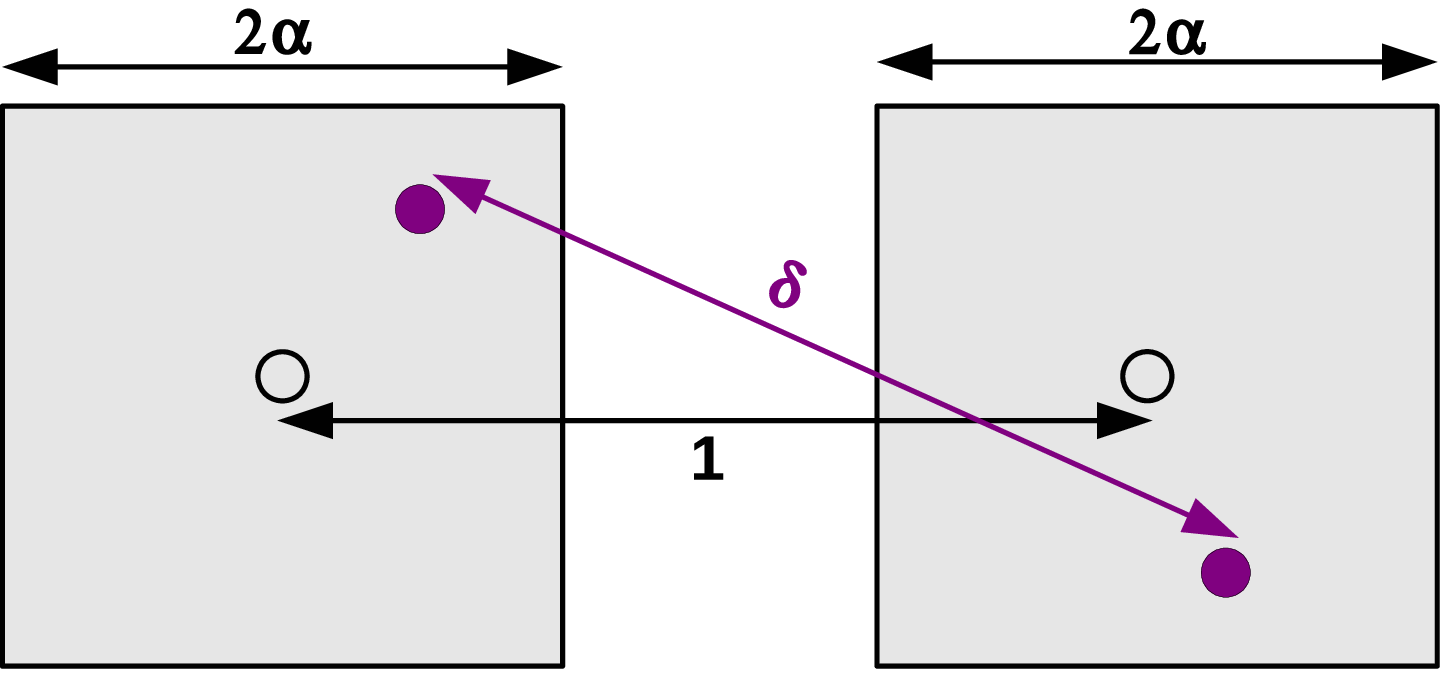}}
\caption{(Color online)Two typical representations of $5 \times 5$ distorted lattices are shown with (a) $\alpha=0.1$ and
(b) $\alpha=0.2$. The intersection points of the dotted lines are the undistorted lattice positions and blue/red circles
represent the distorted positions. A possible configuration of two neighboring lattice points (magenta) is shown in (c).
These two points may take any position within their respective squares (in gray) of length $2\alpha$ centered at
the undistorted positions (open circles). The undistorted distance is taken to be $1$, whereas, their distance
$\delta$ in the distorted lattice may vary between $\delta_m$ and $\delta_M$.}
\label{distlat}
\end{figure*}

In the present work, we introduce a new model of percolation in a distorted square lattice. To begin with, a regular empty
square lattice has been considered. The positions of the sites are then shifted to create a distorted
lattice. The amount of shifts are not same for all the sites but are controlled by a tunable distortion parameter.
The percolation properties are studied for low to moderate distortion with a vision to work on a lattice that
is not perfectly ordered but not too much disorganized either. In section \ref{lattice} we give the details about
the preparation of lattice. The percolation logic on this lattice is explained in section \ref{process}. We present
our results in section \ref{results}. In \ref{pcvar}, we show the variation of percolation threshold and in \ref{univ}
we explore the scaling behavior of the order parameter and calculate the approximate values of the critical
exponents in order to identify the universality class of the present model. Finally we summarize in section \ref{sum}.

\section{The model}
\subsection{The distorted lattice}\label{lattice}
$L \times L$ distorted lattice is created by slightly ruffling the sites of a regular $L \times L$ lattice.
This has been done systematically by setting a distortion parameter $\alpha$ which denotes the maximum amount of
dislocation along $x$ or $y$ axis. Such a lattice can be generated using the following steps:\\
\begin{itemize}
 \item Initially an empty square lattice with equidistant nearest neighbors is considered. The
 lattice constant is set to unity. 
 \item A suitable value for the distortion parameter $\alpha$ is fixed. Since the undistorted distance
 is set to $1$, $\alpha$ may be varied within the range: $0<\alpha<0.5$.
 \item  A lattice site at position $(i,j)$ is chosen. Two random numbers, $r_x$ and $r_y$, are generated
 for $x$ and $y$ direction respectively, in the range $\{-\alpha,\alpha\}$. The position of this site
 is modified to $(X_{ij},Y_{ij})$, where, $X_{ij}=i+r_x$ and $Y_{ij}=j+r_y$.
 \item The above step is repeated for every site of the lattice. A distorted lattice is thereby created.
\end{itemize}
Two suggestive representations of distorted lattice are shown in figure \ref{distlat} for two different values of
$\alpha$. As shown therein, each site can now be located at any point $(i+r_x,j+r_y)$ within a square of length $2\alpha$ with
the undistorted position $(i,j)$ at the center of the square. The distance between any two neighboring sites is denoted
in general by $\delta$. The minimum and maximum limits of $\delta$ are therefore 
\be
\delta_m=1-2\alpha
\ee
and 
\be
\delta_M=\sqrt{(1+2\alpha)^2+(2\alpha)^2}
\ee
respectively. Note that for $\alpha> 0.5$, the lattice is over-distorted: the squares of possible occupancy
of two neighboring sites in figure \ref{distlat}(c) would overlap and the notion of identifying a site with reference
to the regular lattice points $(i,j)$ would lose its meaning. We restrict this study from low to moderate $\alpha$.

\begin{figure*}[t]
\centering
\subfigure[]{\includegraphics[scale=0.45]{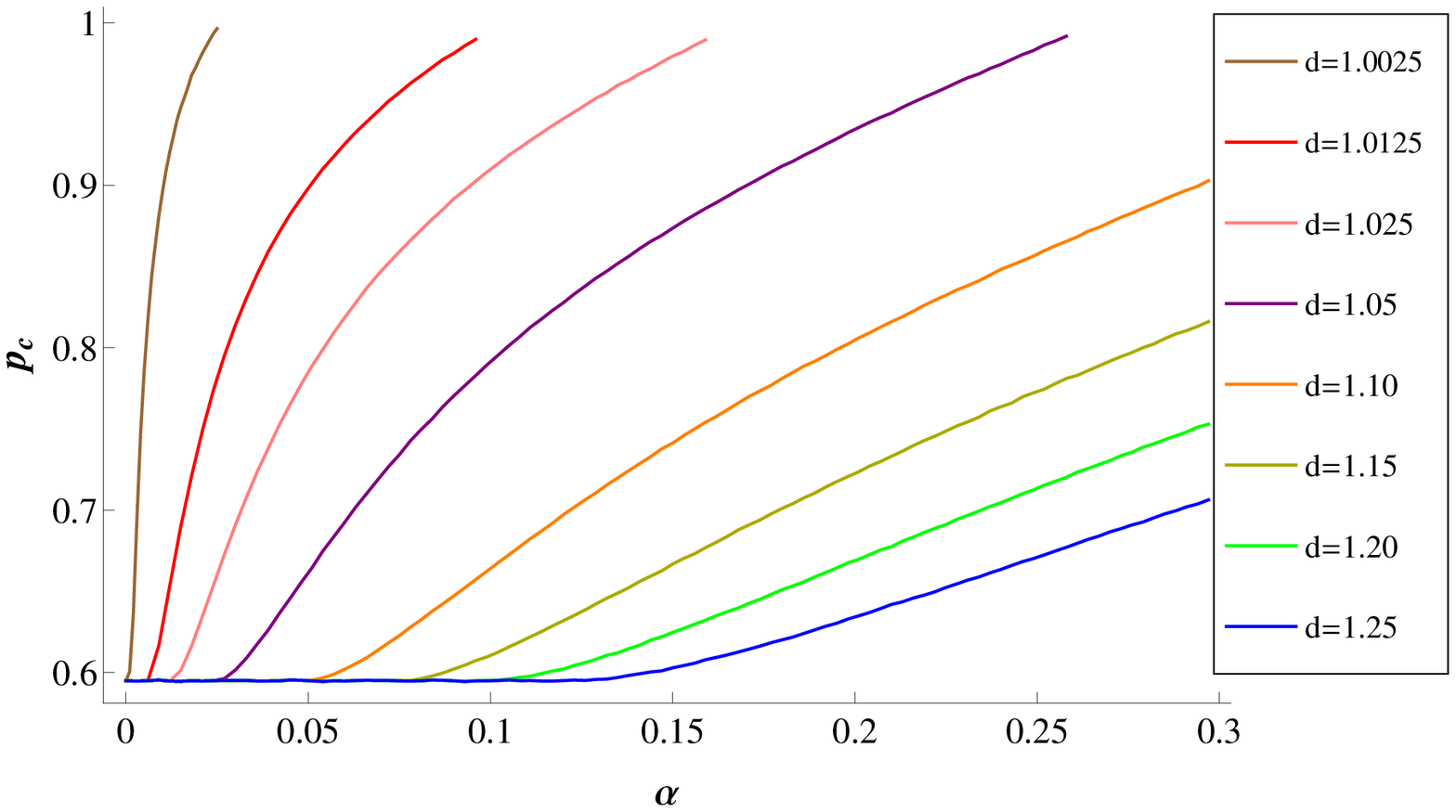}}\hspace{0.3cm}
\subfigure[]{\includegraphics[scale=0.45]{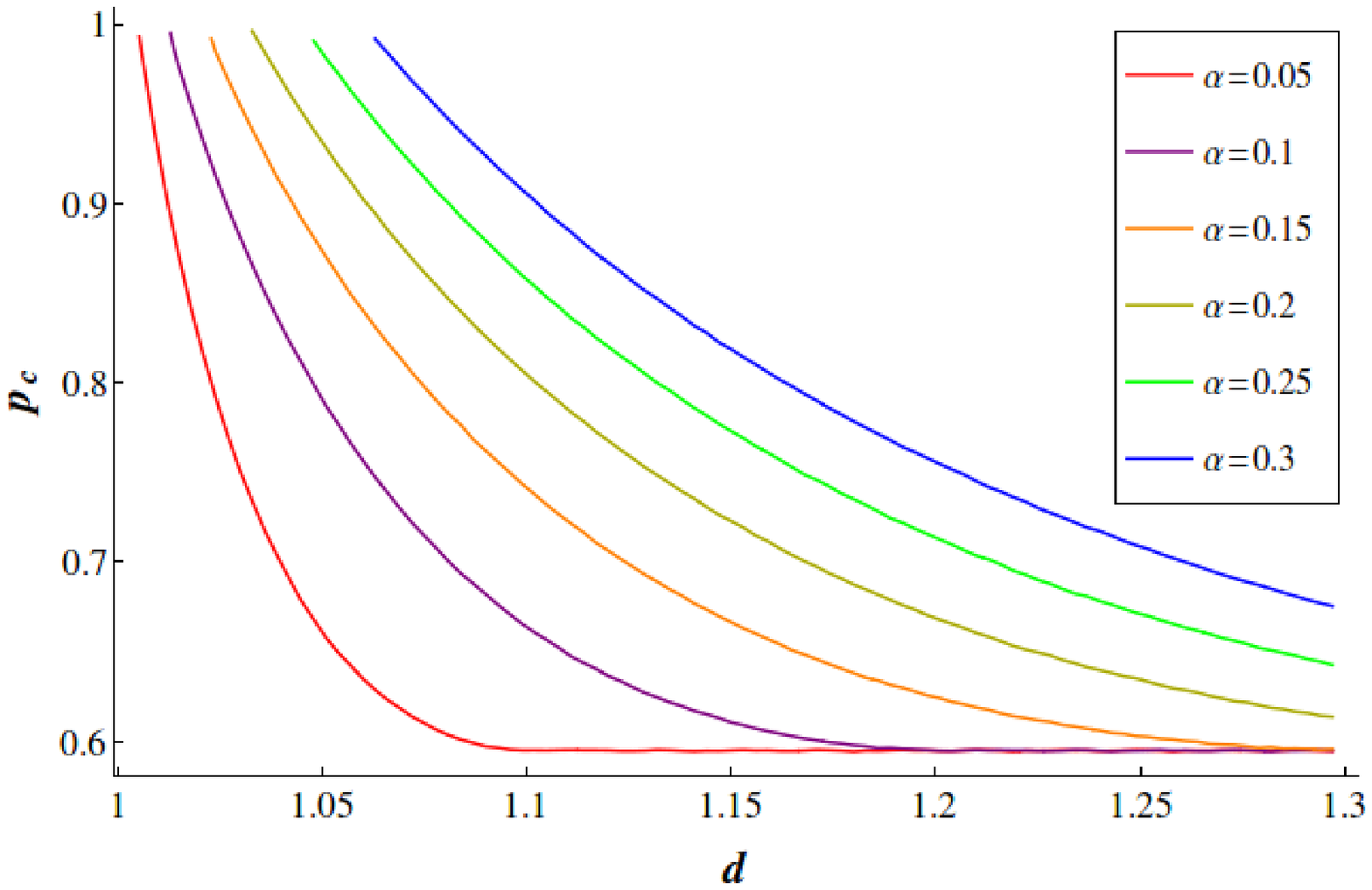}}
 \caption{(Color online) (a) Variation of percolation threshold $p_c$ with distortion $\alpha$ for different
 fixed values of the connection threshold $d$ (each curve corresponds to a fixed value of $d$, in the range $1.0025$
 to $1.25$). Distortion is varied within the range $\alpha=0 - 0.3$. The (blue) curve at the bottom corresponds
 to $d=1.25$ and the one (brown) with a sharp rise is for $d=1.0025$.
 (b) Variation of percolation threshold $p_c$ with the connection threshold $d$ for different
 fixed values of distortion $\alpha$ (each curve corresponds to a fixed value of $\alpha$, in the range $0.05$
 to $0.3$). Connection threshold is varied within the range $d=1.0 - 1.3$. The (blue) curve at the top-right
 corresponds to $\alpha=0.3$ and the left-bottom one (red) is for $\alpha=0.05$. 
 All the plots of both the figures are generated for a system size $L=2^{10}$ and each $p_c$-value is obtained by averaging
 over $200$ independent realizations of the lattice for a given set of values of $\alpha$ and $d$.}
 \label{alpc}
\end{figure*}

\subsection{Percolation process}\label{process}
In usual site percolation, some of the empty sites are occupied randomly corresponding to an occupation
probability $p$. In the present model, the following process has been adopted after generating the distorted empty lattice:
\begin{itemize}
 \item The empty sites of the distorted lattice are occupied randomly as per a given $p$.
 \item A connection threshold ($d$) is set.
 \item Distance ($\delta$) between any pair of occupied neighbors is calculated. These two neighbors are
 considered to be connected if $\delta\le d$; otherwise, the link is broken.
 \item The above step is repeated for each pair of occupied neighbors and the clusters are identified
 accordingly to determine the possibility of percolation.
\end{itemize}

It is clear from the above scheme that even if two nearest neighbors are occupied, they may {\it not} belong
to the same cluster. This is the main difference of the present model with usual site percolation.
The connectivity of any two occupied neighbors depends on $\delta$ (and therefore, on $\alpha$) and on the value
of $d$. Since $\delta_m\le \delta \le \delta_M$, the range of interest for the connection threshold $d$ is also
$\delta_m\le d \le \delta_M$. If $d < \delta_m$, no cluster formation is possible; whereas, for $d > \delta_M$,
every occupied neighbor is connected and the case is similar to usual (undistorted) site percolation. However,
as we shall see in the next section, when $d\le 1$, the system suddenly ceases to percolate even for $p=1.0$,
making the range $\delta_m < d \le 1$ uneventful.

In this work, the cluster identification and numbering has been done by the well known Hoshen-Kopelman (HK)
algorithm \cite{HK}. The connectivity criterion ($\delta\le d$) has been incorporated into the HK algorithm
to appropriately reflect the properties of distorted lattice. We emphasize that this is a controlled site
percolation model and is clearly distinct from bond percolation (where every site is occupied and bonds are
occupied randomly) and site-bond percolation (where both the sites and the bonds between the occupied neighbors
are occupied randomly) models.

\section{results and discussions}\label{results}
\subsection{Variation of percolation threshold}\label{pcvar}
Let us first study the effect of this distortion on the percolation threshold($p_c$). For an undistorted square
lattice, this value is known to be $p_c\approx 0.592746=p_{cu}$ as the lattice size tends to infinity. It is not
hard to realize that $p_c$ depends on the relative strengths of $\alpha$ and $d$ in a distorted lattice.

To demonstrate these dependences
clearly, we first calculate $p_c(\alpha)$ for different fixed values of $d$. Fig. \ref{alpc}(a) shows eight curves,
one each for a value of $d$ ranging between $1.0$ and $1.25$.  All the curves stay at the value $p_{cu}$ as long as
$\alpha$ is low enough, so that $\delta_M<d$. For example, the (blue) curve for $d=1.25$ at the bottom
(see fig. \ref{alpc}(a)) remains at $p_{cu}$ up to an appreciable value of $\alpha$. This is expected since this 
situation is similar to undistorted percolation (even if $\alpha$ is non-zero) due to large value of $d$. The maximum
distance $\delta_M$ between the neighboring points has to exceed the connection threshold $d$ for the manifestation
of any effect of distortion. For lesser values of $d$, percolation threshold is affected by less distortion.
It may be concluded from these plots that when the connection threshold is held fixed, $p_c$ increases with $\alpha$.
This means more distortion makes it more difficult for the system to percolate. This result can be attributed to
the fact that the average distance between two neighboring points increases with $\alpha$ as
$\delta_{min}(\alpha)\approx 1+0.337629\alpha^2$. For lower values of the connection
threshold ($d=1.05, 1.025, 1.0125, 1.0025$ here), $p_c$ reaches very close to $1$ at a certain value of $\alpha$.
As $d$ becomes smaller, this situation occurs with smaller $\alpha$. This indicates that
all the sites need to be occupied to span the lattice. Moreover, if $\alpha$ is further increased the system is no longer
guaranteed to percolate. In fig. \ref{alpc}(a), each $p_c$ is calculated by generating $200$ independent representations
for specific set of $\alpha$ and $d$. For each value of $d$, $p_c$ is shown up to the value of $\alpha$ for which all the $200$
representations percolate. At $d=1.0$, the system ceases to percolate for any non-zero value of $\alpha$. We have also
checked that this situation persists for any $d<1.0$. Thus, for any value of connection threshold which is less than
or equal to the lattice constant (or, the undistorted nearest neighbor distance, here taken to be $1.0$),
the system can not percolate if any distortion, be it very small, is present.

This is an important observation in the context of realistic applications of percolation, particularly since
a perfectly ordered natural system can hardly be found. Consider, for example the simulation of
the forest-fire model. Here, $d$ can be interpreted as the fire-spreading threshold. A distorted array
will make it difficult for the fire to percolate. Moreover, depending on the relative values of $\alpha$ and $d$, 
distortion can even completely stop fire-percolation in a forest where the fire would have definitely percolated
for an undistorted array of same number of trees.

Variation of the percolation threshold($p_c$) with the connection threshold($d$) is shown in fig. \ref{alpc}(b)
for six different values of the distortion parameter ($\alpha$). A higher value of $d$ is expected to favor percolation and this
is evident from fig. \ref{alpc}(b), which shows that $p_c$ decreases with $d$ when $\alpha$ remains fixed. If $d$
is large enough (and $\alpha$ is small enough) to ensure that $d>\delta_M$, the effect of distortion disappears and
consequently, the percolation threshold stays at $p_c=p_{cu}$. In the other extreme, spanning becomes more difficult
with a low connection threshold and beyond a certain value, system can not percolate even after occupying all the
sites. For each $\alpha$, the displayed data start with the minimum value of $d$ which ensures that all the $200$ independent 
realizations do necessarily percolate.

\begin{figure}[h]
\centering
 \subfigure[]{\includegraphics[width=0.45\columnwidth]{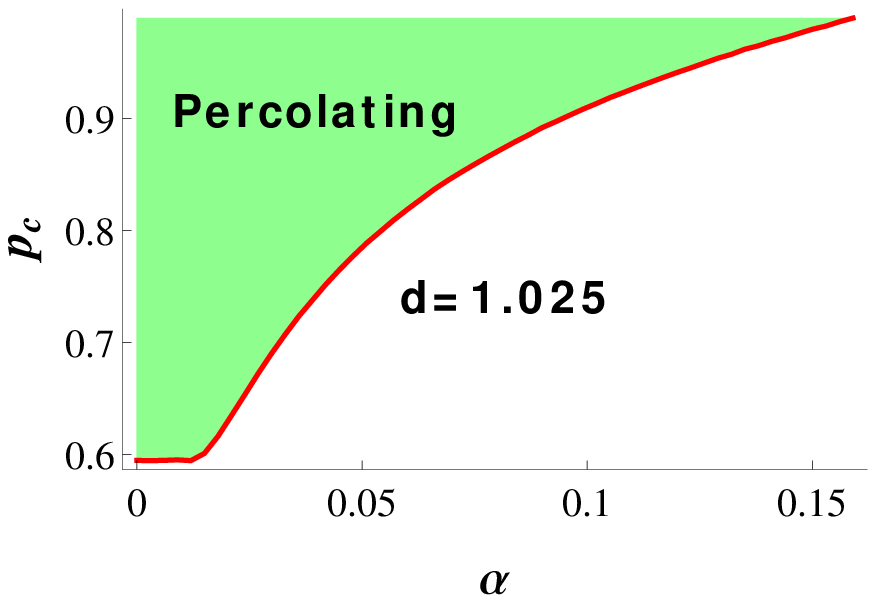}}
 \subfigure[]{\includegraphics[width=0.45\columnwidth]{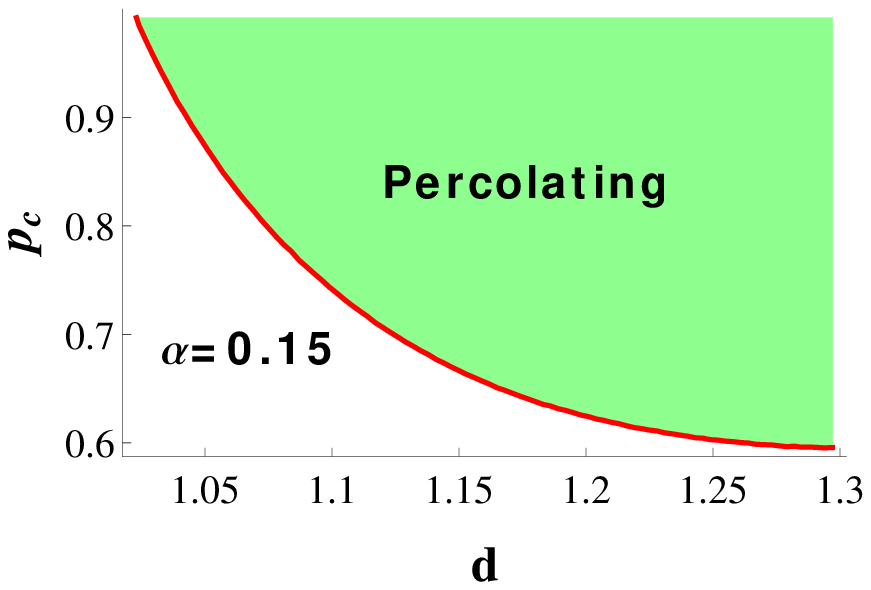}}
 \subfigure[]{\includegraphics[width=0.8\columnwidth]{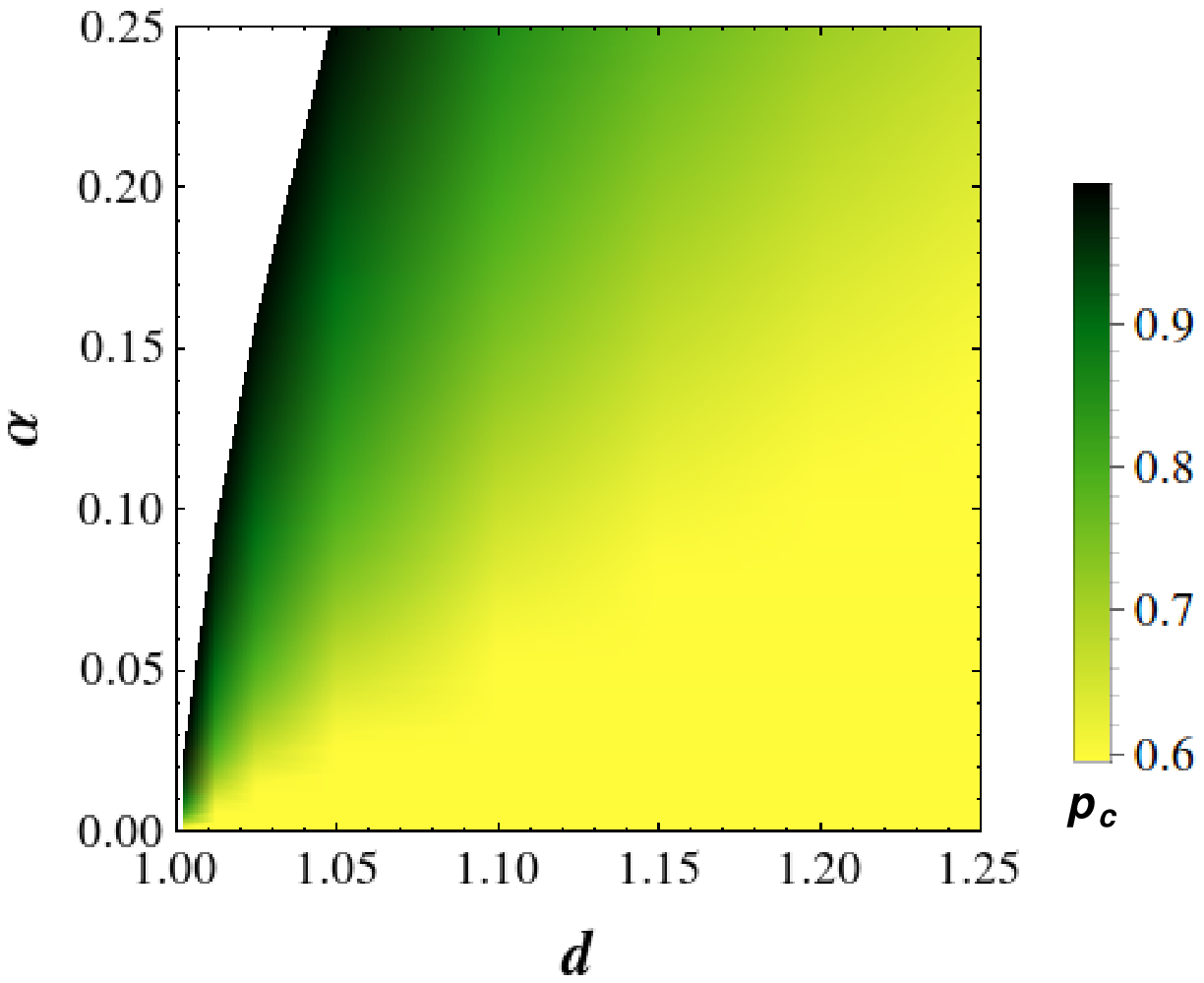}}
 \caption{(Color online) Plots of (a) $p_c(\alpha)$ with $d=1.025$ and (b) $p_c(d)$ with $\alpha=0.15$. The curves
 (and all the other curves in fig. \ref{alpc}(a) and fig. \ref{alpc}(b)) separate the spaces in percolating and
 non-percolating phases. (c) A density plot for $p_c(d,\alpha)$.}
 \label{phase}
\end{figure}

All the plots of fig.\ref{alpc}(a) and fig.\ref{alpc}(b) can be regarded as separation curves between percolating and
non-percolating phases. This is shown in fig.\ref{phase}. For example, if $d=1.025$ and $\alpha=0.05$, spanning
is guaranteed if $85\%$ of the sites are occupied, since the point ($0.05,0.85$) in fig.\ref{phase}(a) falls in
the percolating (green) region. Similarly, if $\alpha=0.15$ and $d=1.05$, a $70\%$ occupancy is not sufficient
for spanning (see fig.\ref{phase}(b)).

The density plot (fig.\ref{phase}(c)) shows the variation of the percolation threshold with both connection
threshold and distortion. Higher $p_c$ is obtained for high $\alpha$ and low $d$. The  blank portion on the left side
indicates that the system can not percolate for those values $\alpha$ and $d$.

\begin{figure*}[t]
\centering
\subfigure[]{\includegraphics[width=\columnwidth]{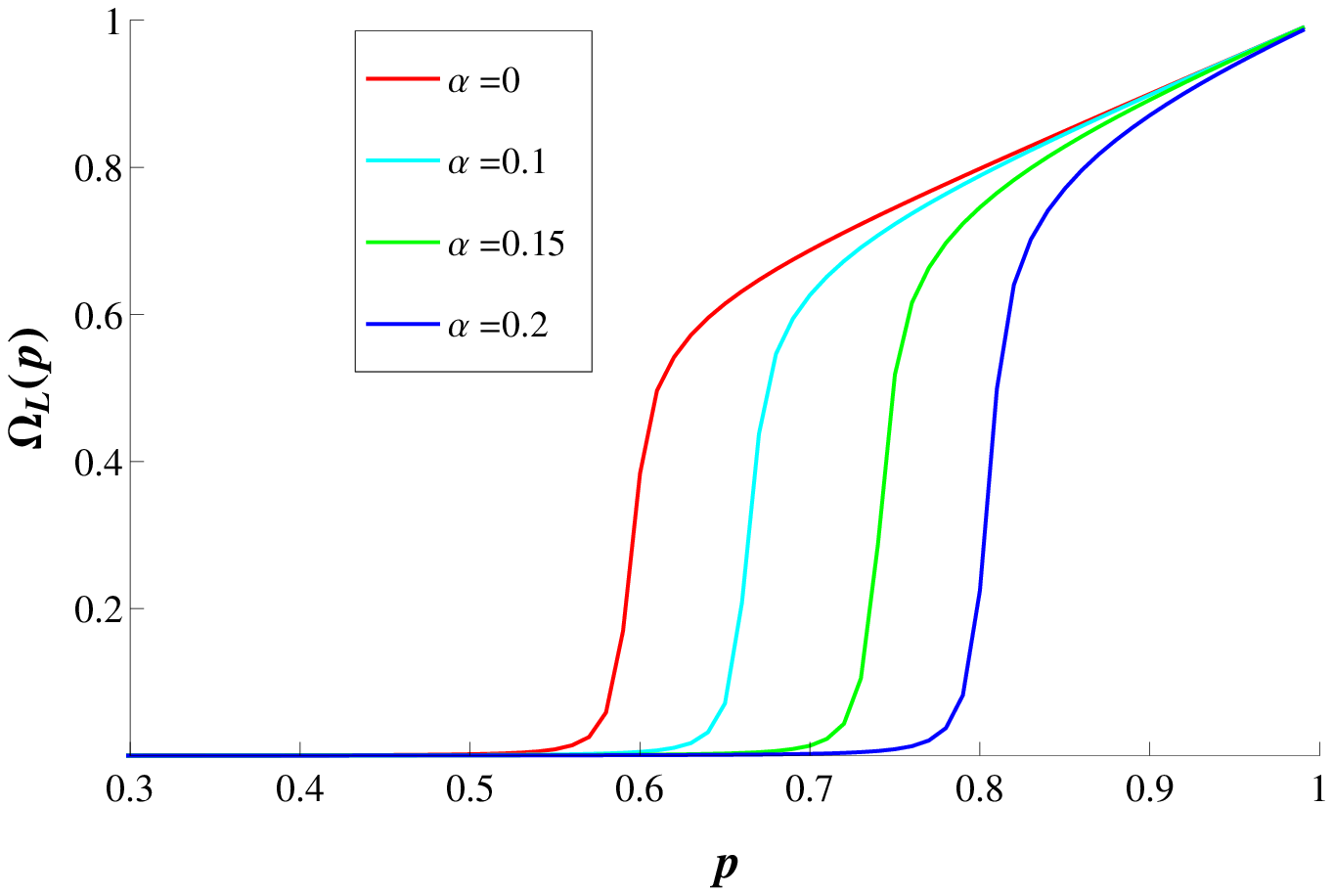}}
 \subfigure[]{\includegraphics[width=\columnwidth]{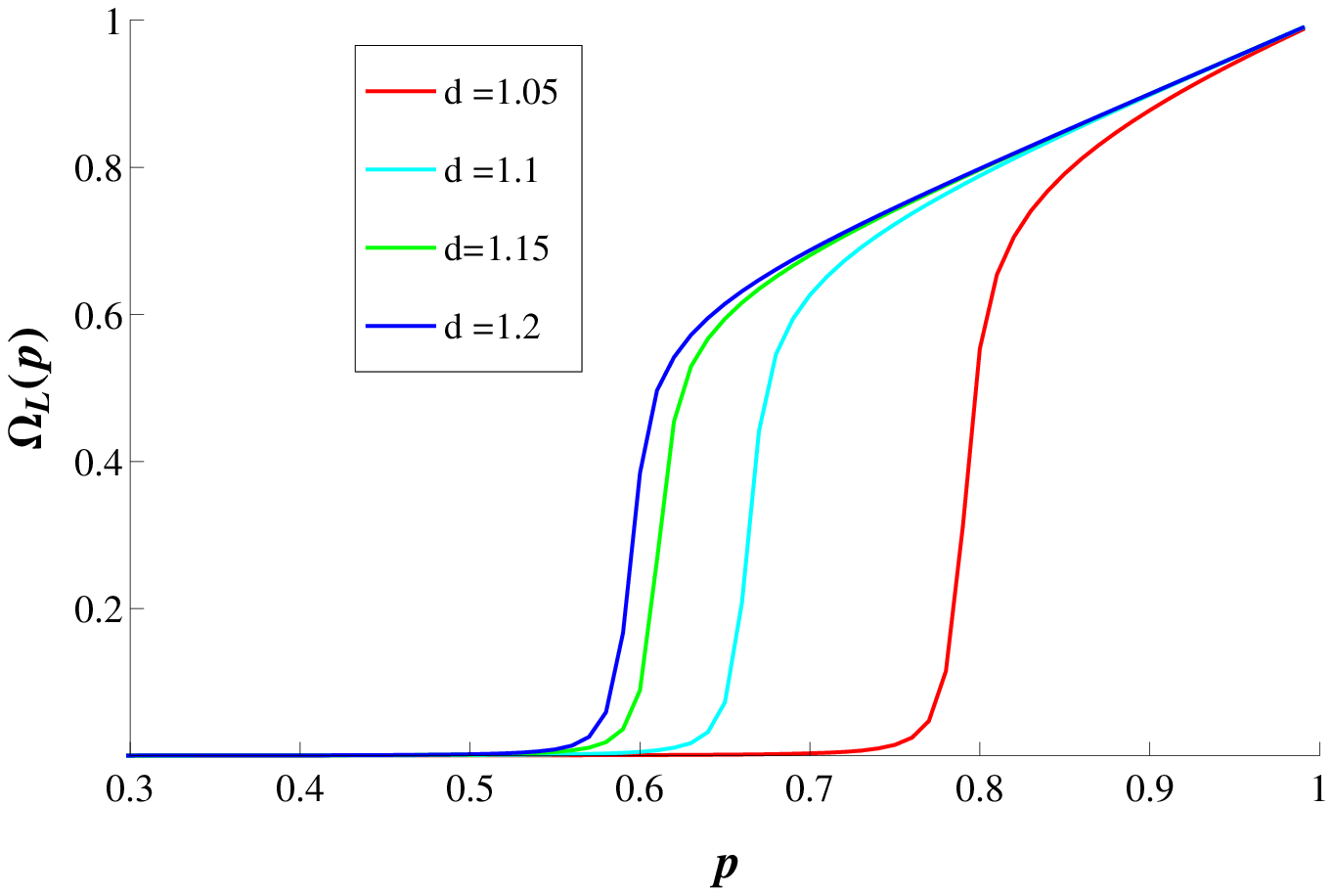}}
 \caption{(Color online) Plots of the order parameter $\Omega_L(p)$ with $p$ (a) for different values of $\alpha$
 keeping $d=1.1$ and (b) for different values of $d$ keeping $\alpha=0.1$. The curves shift towards left as the 
 effect of distortion ceases when $\alpha$ decreases and $d$ increases. Each data point of both the plots are generated
 by averaging over $1000$ independent realizations of the distorted lattice.}
 \label{smax}
\end{figure*}
\subsection{Order parameter and universality class}\label{univ}

\begin{figure*}[t]
\centering
 \subfigure[]{\includegraphics[width=0.95\columnwidth]{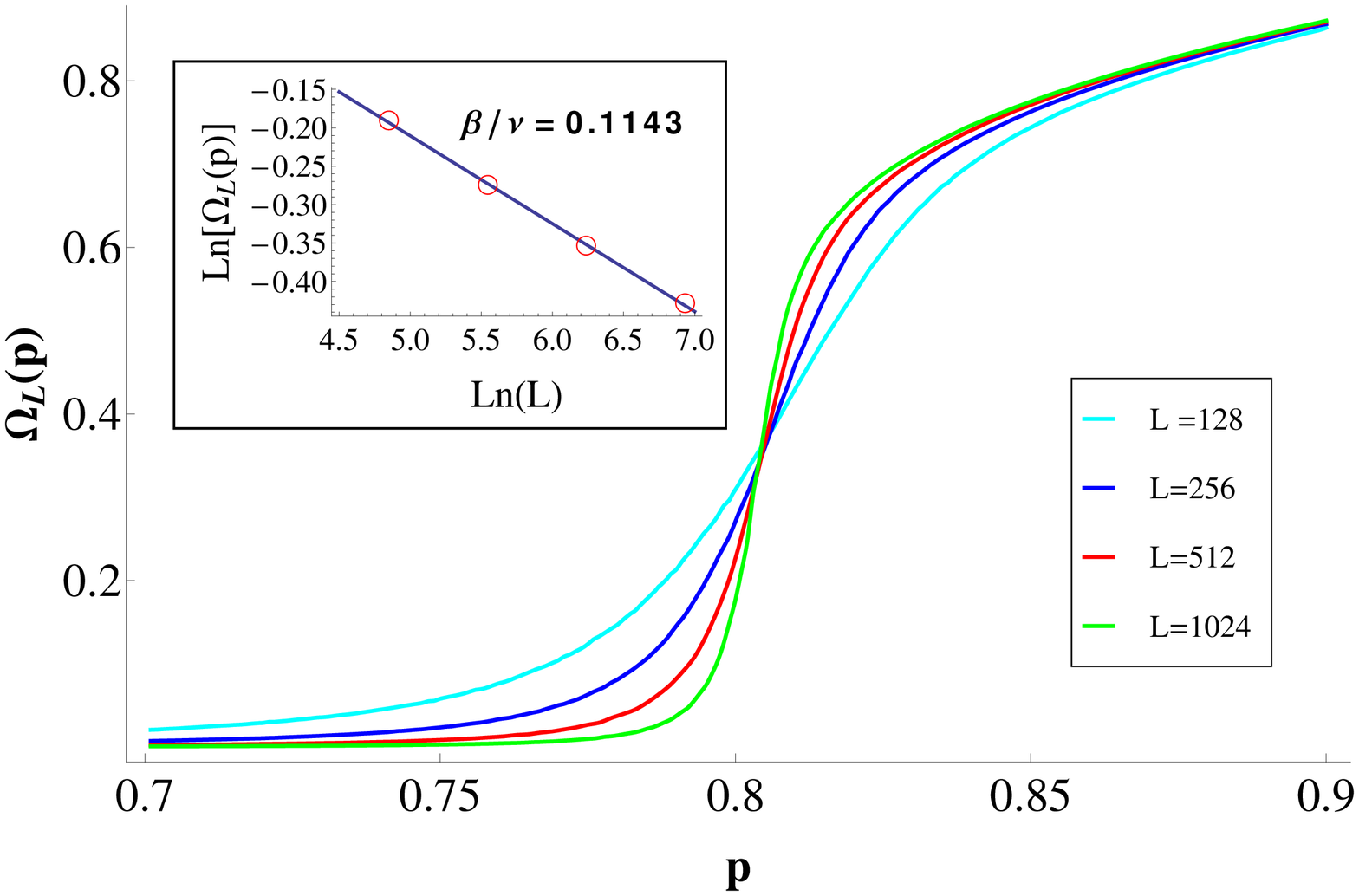}}
\subfigure[]{\includegraphics[width=0.9\columnwidth]{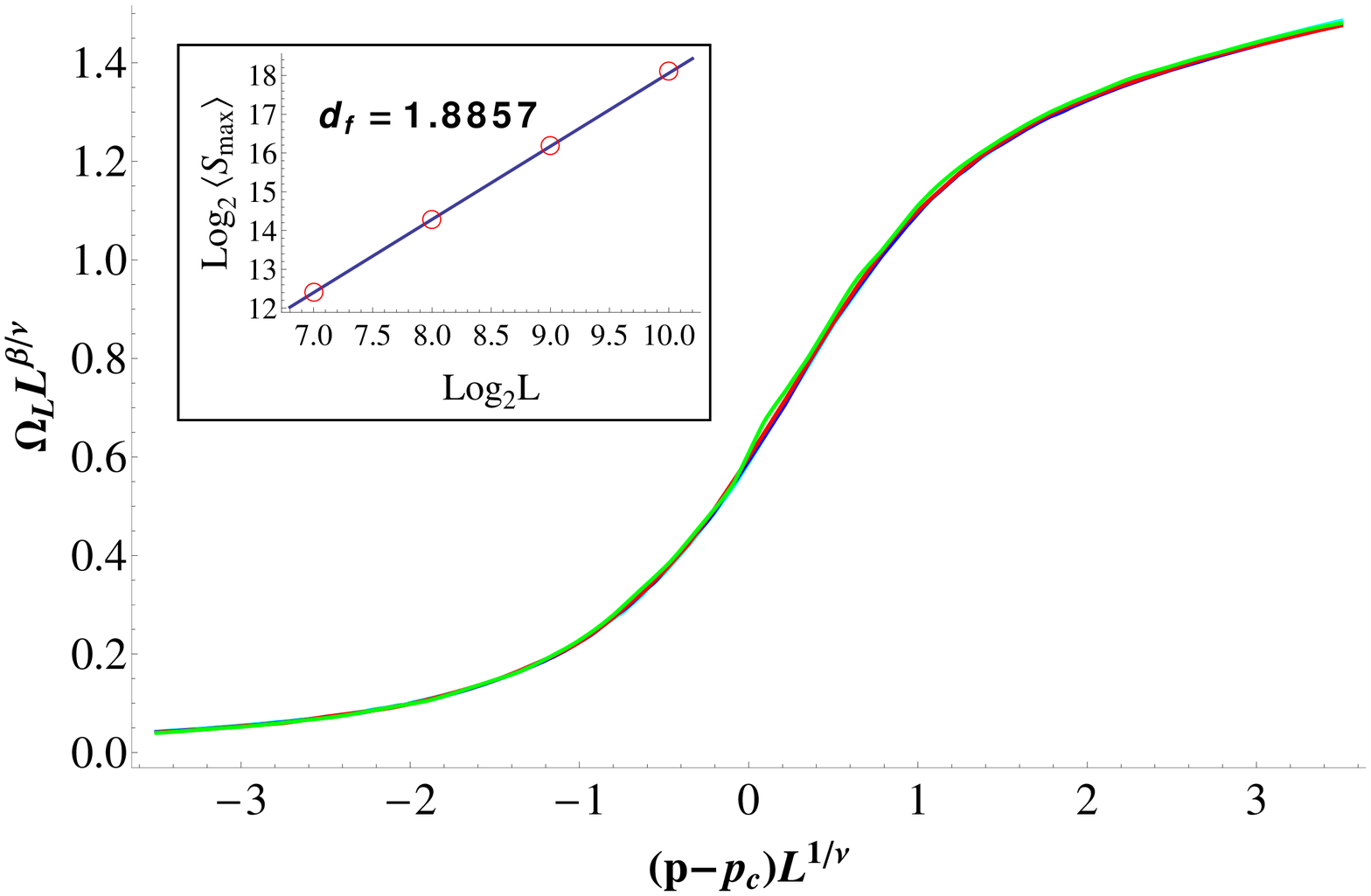}} 
  \caption{(Color online) (a) Plots of $\Omega_L(p)$ for different system sizes $L=2^6,2^7,2^8,2^9,2^{10}$ with
  $\alpha=0.2$ and $d=1.1$ for all the points. [Inset] A log-log plot $\Omega_L(p)$ with $L$ at a particular value
  of $(p-p_c)L^{1/\nu}$ along with a straight line fit of the data gives $\beta/\nu=0.1143$. (b) A nice data
  collapse obtained when $\Omega_L(p) L^{\beta/\nu}$ is plotted against $(p-p_c)L^{1/\nu}$. [Inset] Plot of
  $\log_2\langle S_{max}\rangle$ at $p=p_c$ with $\log_2L$ and a straight line fit gives $d_f=1.8857$, consistent with
  $d_f=2-\beta/\nu$.}
 \label{scaling}
\end{figure*}

The percolation order parameter is usually defined as \begin{equation}
                                                       \Omega_L(p)=\langle S_{max}\rangle/L^2, 
                                                      \end{equation}

where $S_{max}$ stands for the number of occupied sites in the largest cluster and $\langle\rangle$ denotes
configuration-average. In order to calculate $\Omega_L(p)$ for an $L\times L$ system , one needs to occupy the sites as per
the occupation probability $p$, count the number of sites in the largest cluster, average over many such configurations
and finally divide it by $L^2$. In fig. \ref{smax}(a), $\Omega_L(p)$ is plotted for four values of the distortion
parameter $\alpha=0,0.1,0.15,0.2$ from left to right for a fixed $d=1.1$. As the effect of distortion diminishes,
the curve shifts towards left. Similar situation is displayed in fig. \ref{smax}(b); here the influence of distortion
is reduced by increment in connection threshold ($d$), although $\alpha=0.1$ remains unchanged.
In fig. \ref{smax}(b), the two curves on the left are very close to each other.
This indicates that the undistorted scenario is being approached and the curves with still higher $d$ would be identical.

Figure \ref{scaling}(a) shows $\Omega_L(p)$ for different system sizes: $L=128,256,512,1024$, with the steepest one
corresponding to the largest $L$. The parameters of $\alpha=0.2$ and $d=1.1$ have been held fixed. For this set of
values we find
$p_c=0.8025$. Using the value of the critical exponent $\nu=0.75$ for standard percolation, when the horizontal
axis is scaled as $(p-p_c)L^{1/\nu}$, the curves of $\Omega_L(p)$ separate from each other
with the $L=128$ curve being on top and the $L=1024$ one at the bottom. The values of $\Omega_L(p)$ are collected
at a fixed value of $(p-p_c)L^{1/\nu}$ and using these values a plot of $\ln \Omega_L(p)$ with $\ln L$ is generated
(see inset of figure \ref{scaling}(a)). The slope of the straight line fit of this data gives $\beta/\nu=0.1143$.
As known from standard percolation criticality, if the vertical axis is now scaled as $\Omega_L(p) L^{\beta/\nu}$ and
plotted against $(p-p_c)L^{1/\nu}$, a data collapse should be obtained. We do get a nice data collapse in
figure \ref{scaling}(b) using the obtained value of $\beta/\nu$.

It is known that the percolating cluster at $p=p_c$ is a fractal whose fractal-dimension $d_f$ is given by
\begin{equation}
 \langle S_{max}\rangle \sim L^{d_f}.
\label{dfeq}
\end{equation}

We calculate $\langle S_{max}\rangle$
for $L=2^7,2^8,2^9,2^{10}$ with $\alpha=0.2$ and $d=1.1$ at $p=0.8025=p_c$. Eq. \ref{dfeq} suggests that a log-log plot
of $\langle S_{max}\rangle$ and $L$ would fit into a straight line with its slope$=d_f$. In the inset of figure
\ref{scaling}(b), we show this plot along with a linear fit that gives $d_f=1.8857$. This confirms the well known
relation between the fractal dimension of the percolating cluster at $p_c$ and the relevant critical exponents in
two dimensional systems
\begin{equation}
 d_f=2-\beta/\nu.
\end{equation}

The whole process has been repeated for two other sets of values of $\alpha$ and $d$ (plots are not shown as they look
very similar to those given in figure \ref{scaling}). For $\alpha=0.1, d=1.1$, we get
$p_c=0.6617, \beta/\nu=0.1150$ and for $\alpha=0.2, d=1.2$, we get $p_c=0.6665, \beta/\nu=0.1165$. These values
should be compared to those for the ordinary percolation, for which, $\beta/\nu=0.1012$ and $d_f=1.8958$.

These results tend to indicate that the present variant of percolation in a distorted
lattice may belong to the same universality class as the ordinary percolation. It has to be understood that within
the scope of the present work, the values of the exponents,
percolation threshold and fractal dimension are suggestive. More precise values, rigorous analysis
on their dependence on $\alpha$ and $d$ and a conclusive decision on universality class require further detailed
study and extensive numerical calculation involving averages over larger number of configurations with
(preferably) bigger lattice sizes. We plan for such a detailed study in our future endeavor.

\section{summary}\label{sum}
To summarize, we have proposed a new model of percolation in which the empty sites of a regular square lattice
are distorted. Distortion is incorporated into the system through a parameter $\alpha$, which randomly shifts
the position of each site within a square of length $2\alpha$ centered at the regular location of the site. 
Thus the nearest neighbor distance may be more, equal or less than that of the undistorted
lattice. Two adjacent occupied sites are considered connected only when their distance is less
than a predefined value, called connection threshold $d$. In a Monte-Carlo study via HK algorithm, we find that 
spanning becomes difficult for higher values of $\alpha$ and lower values of $d$. The value of the percolation
threshold $p_c$ depends on its interplay with the two parameters $\alpha$ and $d$ and varies within $p_{cu}\le p_c\le 1$.
Interestingly, if $d$ is less or equal to its value in the undistorted lattice (we take this value to be unity),
the system fails to percolate with any non-zero $\alpha$ (be it very small, meaning slight distortion) even when
all the sites are occupied. From the obtained values of the fractal dimension of spanning cluster at $p_c$ and the critical
exponents related to it, we predict with caution that this model may belong to the same universality class as usual
site percolation.
\begin{acknowledgments}
The computation facilities availed at the Department of Physics, University of Gour Banga, Malda is gratefully
acknowledged. One of us (AS) would like to thank Raja Paul of IACS, Kolkata for an illuminating discussion. 
\end{acknowledgments}


%

\begin{thebibliography}{34}%
\makeatletter
\providecommand \@ifxundefined [1]{%
 \@ifx{#1\undefined}
}%
\providecommand \@ifnum [1]{%
 \ifnum #1\expandafter \@firstoftwo
 \else \expandafter \@secondoftwo
 \fi
}%
\providecommand \@ifx [1]{%
 \ifx #1\expandafter \@firstoftwo
 \else \expandafter \@secondoftwo
 \fi
}%
\providecommand \natexlab [1]{#1}%
\providecommand \enquote  [1]{``#1''}%
\providecommand \bibnamefont  [1]{#1}%
\providecommand \bibfnamefont [1]{#1}%
\providecommand \citenamefont [1]{#1}%
\providecommand \href@noop [0]{\@secondoftwo}%
\providecommand \href [0]{\begingroup \@sanitize@url \@href}%
\providecommand \@href[1]{\@@startlink{#1}\@@href}%
\providecommand \@@href[1]{\endgroup#1\@@endlink}%
\providecommand \@sanitize@url [0]{\catcode `\\12\catcode `\$12\catcode
  `\&12\catcode `\#12\catcode `\^12\catcode `\_12\catcode `\%12\relax}%
\providecommand \@@startlink[1]{}%
\providecommand \@@endlink[0]{}%
\providecommand \url  [0]{\begingroup\@sanitize@url \@url }%
\providecommand \@url [1]{\endgroup\@href {#1}{\urlprefix }}%
\providecommand \urlprefix  [0]{URL }%
\providecommand \Eprint [0]{\href }%
\providecommand \doibase [0]{http://dx.doi.org/}%
\providecommand \selectlanguage [0]{\@gobble}%
\providecommand \bibinfo  [0]{\@secondoftwo}%
\providecommand \bibfield  [0]{\@secondoftwo}%
\providecommand \translation [1]{[#1]}%
\providecommand \BibitemOpen [0]{}%
\providecommand \bibitemStop [0]{}%
\providecommand \bibitemNoStop [0]{.\EOS\space}%
\providecommand \EOS [0]{\spacefactor3000\relax}%
\providecommand \BibitemShut  [1]{\csname bibitem#1\endcsname}%
\let\auto@bib@innerbib\@empty
\bibitem [{\citenamefont {Stauffer}\ and\ \citenamefont
  {Aharony}(1994)}]{book}%
  \BibitemOpen
  \bibfield  {author} {\bibinfo {author} {\bibfnamefont {D.}~\bibnamefont
  {Stauffer}}\ and\ \bibinfo {author} {\bibfnamefont {A.}~\bibnamefont
  {Aharony}},\ }\href@noop {} {\emph {\bibinfo {title} {Introduction to
  Percolation Theory, 2nd ed.}}}\ (\bibinfo  {publisher} {Taylor and Francis},\
  \bibinfo {address} {London},\ \bibinfo {year} {1994})\BibitemShut {NoStop}%
\bibitem [{\citenamefont {Saberi}(2015)}]{review}%
  \BibitemOpen
  \bibfield  {author} {\bibinfo {author} {\bibfnamefont {A.~A.}\ \bibnamefont
  {Saberi}},\ }\href@noop {} {\bibfield  {journal} {\bibinfo  {journal} {Phys.
  Rep.}\ }\textbf {\bibinfo {volume} {578}},\ \bibinfo {pages} {1} (\bibinfo
  {year} {2015})}\BibitemShut {NoStop}%
\bibitem [{\citenamefont {Broadbent}\ and\ \citenamefont
  {Hammersley}(1957)}]{Broadbent}%
  \BibitemOpen
  \bibfield  {author} {\bibinfo {author} {\bibfnamefont {S.~R.}\ \bibnamefont
  {Broadbent}}\ and\ \bibinfo {author} {\bibfnamefont {J.~M.}\ \bibnamefont
  {Hammersley}},\ }\href@noop {} {\bibfield  {journal} {\bibinfo  {journal}
  {Mathematical Proceedings of the Cambridge Philosophical Society}\ }\textbf
  {\bibinfo {volume} {53}},\ \bibinfo {pages} {629} (\bibinfo {year}
  {1957})}\BibitemShut {NoStop}%
\bibitem [{\citenamefont {Isichenko}(1992)}]{Isichenko}%
  \BibitemOpen
  \bibfield  {author} {\bibinfo {author} {\bibfnamefont {M.~B.}\ \bibnamefont
  {Isichenko}},\ }\href@noop {} {\bibfield  {journal} {\bibinfo  {journal}
  {Rev. Mod. Phys.}\ }\textbf {\bibinfo {volume} {64}},\ \bibinfo {pages} {961}
  (\bibinfo {year} {1992})}\BibitemShut {NoStop}%
\bibitem [{\citenamefont {Ball}\ \emph {et~al.}(1994)\citenamefont {Ball},
  \citenamefont {Phillips}, \citenamefont {Callahan},\ and\ \citenamefont
  {Sauerbrey}}]{Ball}%
  \BibitemOpen
  \bibfield  {author} {\bibinfo {author} {\bibfnamefont {Z.}~\bibnamefont
  {Ball}}, \bibinfo {author} {\bibfnamefont {H.~M.}\ \bibnamefont {Phillips}},
  \bibinfo {author} {\bibfnamefont {D.~L.}\ \bibnamefont {Callahan}}, \ and\
  \bibinfo {author} {\bibfnamefont {R.}~\bibnamefont {Sauerbrey}},\ }\href@noop
  {} {\bibfield  {journal} {\bibinfo  {journal} {Phys. Rev. Lett.}\ }\textbf
  {\bibinfo {volume} {73}},\ \bibinfo {pages} {2099} (\bibinfo {year}
  {1994})}\BibitemShut {NoStop}%
\bibitem [{\citenamefont {Dotsenko}\ \emph {et~al.}(1993)\citenamefont
  {Dotsenko}, \citenamefont {Windey}, \citenamefont {Harris}, \citenamefont
  {Marinari}, \citenamefont {Martinec},\ and\ \citenamefont
  {Picco}}]{Dotsenko}%
  \BibitemOpen
  \bibfield  {author} {\bibinfo {author} {\bibfnamefont {V.~S.}\ \bibnamefont
  {Dotsenko}}, \bibinfo {author} {\bibfnamefont {P.}~\bibnamefont {Windey}},
  \bibinfo {author} {\bibfnamefont {G.}~\bibnamefont {Harris}}, \bibinfo
  {author} {\bibfnamefont {E.}~\bibnamefont {Marinari}}, \bibinfo {author}
  {\bibfnamefont {E.}~\bibnamefont {Martinec}}, \ and\ \bibinfo {author}
  {\bibfnamefont {M.}~\bibnamefont {Picco}},\ }\href@noop {} {\bibfield
  {journal} {\bibinfo  {journal} {Phys. Rev. Lett.}\ }\textbf {\bibinfo
  {volume} {71}},\ \bibinfo {pages} {811} (\bibinfo {year} {1993})}\BibitemShut
  {NoStop}%
\bibitem [{\citenamefont {Gruzberg}\ \emph {et~al.}(1999)\citenamefont
  {Gruzberg}, \citenamefont {Ludwig},\ and\ \citenamefont {Read}}]{Gruzberg}%
  \BibitemOpen
  \bibfield  {author} {\bibinfo {author} {\bibfnamefont {I.~A.}\ \bibnamefont
  {Gruzberg}}, \bibinfo {author} {\bibfnamefont {A.~W.~W.}\ \bibnamefont
  {Ludwig}}, \ and\ \bibinfo {author} {\bibfnamefont {N.}~\bibnamefont
  {Read}},\ }\href@noop {} {\bibfield  {journal} {\bibinfo  {journal} {Phys.
  Rev. Lett.}\ }\textbf {\bibinfo {volume} {82}},\ \bibinfo {pages} {4524}
  (\bibinfo {year} {1999})}\BibitemShut {NoStop}%
\bibitem [{\citenamefont {Saberi}(2010)}]{Saberi}%
  \BibitemOpen
  \bibfield  {author} {\bibinfo {author} {\bibfnamefont {A.~A.}\ \bibnamefont
  {Saberi}},\ }\href@noop {} {\bibfield  {journal} {\bibinfo  {journal} {Appl.
  Phys. Lett.}\ }\textbf {\bibinfo {volume} {97}},\ \bibinfo {pages} {154102}
  (\bibinfo {year} {2010})}\BibitemShut {NoStop}%
\bibitem [{\citenamefont {{D’Souza}}\ and\ \citenamefont
  {Mitzenmacher}(2010)}]{Dsouza}%
  \BibitemOpen
  \bibfield  {author} {\bibinfo {author} {\bibfnamefont {R.~M.}\ \bibnamefont
  {{D’Souza}}}\ and\ \bibinfo {author} {\bibfnamefont {M.}~\bibnamefont
  {Mitzenmacher}},\ }\href@noop {} {\bibfield  {journal} {\bibinfo  {journal}
  {Phys. Rev. Lett.}\ }\textbf {\bibinfo {volume} {104}},\ \bibinfo {pages}
  {195702} (\bibinfo {year} {2010})}\BibitemShut {NoStop}%
\bibitem [{\citenamefont {Derenyi}\ \emph {et~al.}(2005)\citenamefont
  {Derenyi}, \citenamefont {Palla},\ and\ \citenamefont {Vicsek}}]{Derenyi}%
  \BibitemOpen
  \bibfield  {author} {\bibinfo {author} {\bibfnamefont {I.}~\bibnamefont
  {Derenyi}}, \bibinfo {author} {\bibfnamefont {G.}~\bibnamefont {Palla}}, \
  and\ \bibinfo {author} {\bibfnamefont {T.}~\bibnamefont {Vicsek}},\
  }\href@noop {} {\bibfield  {journal} {\bibinfo  {journal} {Phys. Rev. Lett.}\
  }\textbf {\bibinfo {volume} {94}},\ \bibinfo {pages} {160202} (\bibinfo
  {year} {2005})}\BibitemShut {NoStop}%
\bibitem [{\citenamefont {Callaway}\ \emph {et~al.}(2000)\citenamefont
  {Callaway}, \citenamefont {Newman}, \citenamefont {Strogatz},\ and\
  \citenamefont {Watts}}]{Callaway}%
  \BibitemOpen
  \bibfield  {author} {\bibinfo {author} {\bibfnamefont {D.~S.}\ \bibnamefont
  {Callaway}}, \bibinfo {author} {\bibfnamefont {M.~E.~J.}\ \bibnamefont
  {Newman}}, \bibinfo {author} {\bibfnamefont {S.~H.}\ \bibnamefont
  {Strogatz}}, \ and\ \bibinfo {author} {\bibfnamefont {D.~J.}\ \bibnamefont
  {Watts}},\ }\href@noop {} {\bibfield  {journal} {\bibinfo  {journal} {Phys.
  Rev. Lett.}\ }\textbf {\bibinfo {volume} {85}},\ \bibinfo {pages} {5468}
  (\bibinfo {year} {2000})}\BibitemShut {NoStop}%
\bibitem [{\citenamefont {Kalisky}\ and\ \citenamefont
  {Cohen}(2006)}]{Kalisky}%
  \BibitemOpen
  \bibfield  {author} {\bibinfo {author} {\bibfnamefont {T.}~\bibnamefont
  {Kalisky}}\ and\ \bibinfo {author} {\bibfnamefont {R.}~\bibnamefont
  {Cohen}},\ }\href@noop {} {\bibfield  {journal} {\bibinfo  {journal} {Phys.
  Rev. E}\ }\textbf {\bibinfo {volume} {73}},\ \bibinfo {pages} {035101(R)}
  (\bibinfo {year} {2006})}\BibitemShut {NoStop}%
\bibitem [{\citenamefont {Coniglio}\ \emph {et~al.}(1979)\citenamefont
  {Coniglio}, \citenamefont {Stanley},\ and\ \citenamefont {Klein}}]{Coniglio}%
  \BibitemOpen
  \bibfield  {author} {\bibinfo {author} {\bibfnamefont {A.}~\bibnamefont
  {Coniglio}}, \bibinfo {author} {\bibfnamefont {H.~E.}\ \bibnamefont
  {Stanley}}, \ and\ \bibinfo {author} {\bibfnamefont {W.}~\bibnamefont
  {Klein}},\ }\href@noop {} {\bibfield  {journal} {\bibinfo  {journal} {Phys.
  Rev. Lett.}\ }\textbf {\bibinfo {volume} {42}},\ \bibinfo {pages} {518}
  (\bibinfo {year} {1979})}\BibitemShut {NoStop}%
\bibitem [{\citenamefont {Anekal}\ \emph {et~al.}(2006)\citenamefont {Anekal},
  \citenamefont {Bahukudumbi},\ and\ \citenamefont {Bevan}}]{Anekal}%
  \BibitemOpen
  \bibfield  {author} {\bibinfo {author} {\bibfnamefont {S.~G.}\ \bibnamefont
  {Anekal}}, \bibinfo {author} {\bibfnamefont {P.}~\bibnamefont {Bahukudumbi}},
  \ and\ \bibinfo {author} {\bibfnamefont {M.~A.}\ \bibnamefont {Bevan}},\
  }\href@noop {} {\bibfield  {journal} {\bibinfo  {journal} {Phys. Rev. E}\
  }\textbf {\bibinfo {volume} {73}},\ \bibinfo {pages} {020403} (\bibinfo
  {year} {2006})}\BibitemShut {NoStop}%
\bibitem [{\citenamefont {Gnan}\ \emph {et~al.}(2014)\citenamefont {Gnan},
  \citenamefont {Zaccarelli},\ and\ \citenamefont {Sciortino}}]{Gnan}%
  \BibitemOpen
  \bibfield  {author} {\bibinfo {author} {\bibfnamefont {N.}~\bibnamefont
  {Gnan}}, \bibinfo {author} {\bibfnamefont {E.}~\bibnamefont {Zaccarelli}}, \
  and\ \bibinfo {author} {\bibfnamefont {F.}~\bibnamefont {Sciortino}},\
  }\href@noop {} {\bibfield  {journal} {\bibinfo  {journal} {Nature
  Communications}\ }\textbf {\bibinfo {volume} {5}},\ \bibinfo {pages} {3267}
  (\bibinfo {year} {2014})}\BibitemShut {NoStop}%
\bibitem [{\citenamefont {King}\ \emph {et~al.}(1999)\citenamefont {King},
  \citenamefont {Buldyrev}, \citenamefont {Dokholyan}, \citenamefont {Havlin},
  \citenamefont {Lee}, \citenamefont {Paul},\ and\ \citenamefont
  {Stanley}}]{King}%
  \BibitemOpen
  \bibfield  {author} {\bibinfo {author} {\bibfnamefont {P.~R.}\ \bibnamefont
  {King}}, \bibinfo {author} {\bibfnamefont {S.~V.}\ \bibnamefont {Buldyrev}},
  \bibinfo {author} {\bibfnamefont {N.~V.}\ \bibnamefont {Dokholyan}}, \bibinfo
  {author} {\bibfnamefont {S.}~\bibnamefont {Havlin}}, \bibinfo {author}
  {\bibfnamefont {Y.}~\bibnamefont {Lee}}, \bibinfo {author} {\bibfnamefont
  {G.}~\bibnamefont {Paul}}, \ and\ \bibinfo {author} {\bibfnamefont {H.~E.}\
  \bibnamefont {Stanley}},\ }\href@noop {} {\bibfield  {journal} {\bibinfo
  {journal} {Physica A}\ }\textbf {\bibinfo {volume} {274}},\ \bibinfo {pages}
  {60} (\bibinfo {year} {1999})}\BibitemShut {NoStop}%
\bibitem [{\citenamefont {Sapoval}\ \emph {et~al.}(2004)\citenamefont
  {Sapoval}, \citenamefont {Baldassarri},\ and\ \citenamefont
  {Gabrielli}}]{Sapoval}%
  \BibitemOpen
  \bibfield  {author} {\bibinfo {author} {\bibfnamefont {B.}~\bibnamefont
  {Sapoval}}, \bibinfo {author} {\bibfnamefont {A.}~\bibnamefont
  {Baldassarri}}, \ and\ \bibinfo {author} {\bibfnamefont {A.}~\bibnamefont
  {Gabrielli}},\ }\href@noop {} {\bibfield  {journal} {\bibinfo  {journal}
  {Phys. Rev. Lett.}\ }\textbf {\bibinfo {volume} {93}},\ \bibinfo {pages}
  {098501} (\bibinfo {year} {2004})}\BibitemShut {NoStop}%
\bibitem [{\citenamefont {Saberi}(2013)}]{Saberi_coast}%
  \BibitemOpen
  \bibfield  {author} {\bibinfo {author} {\bibfnamefont {A.~A.}\ \bibnamefont
  {Saberi}},\ }\href@noop {} {\bibfield  {journal} {\bibinfo  {journal} {Phys.
  Rev. Lett.}\ }\textbf {\bibinfo {volume} {110}},\ \bibinfo {pages} {178501}
  (\bibinfo {year} {2013})}\BibitemShut {NoStop}%
\bibitem [{\citenamefont {Albano}(1995)}]{Albano}%
  \BibitemOpen
  \bibfield  {author} {\bibinfo {author} {\bibfnamefont {E.~V.}\ \bibnamefont
  {Albano}},\ }\href@noop {} {\bibfield  {journal} {\bibinfo  {journal}
  {Physica A}\ }\textbf {\bibinfo {volume} {216}},\ \bibinfo {pages} {213}
  (\bibinfo {year} {1995})}\BibitemShut {NoStop}%
\bibitem [{\citenamefont {Albano}(1994)}]{Albano2}%
  \BibitemOpen
  \bibfield  {author} {\bibinfo {author} {\bibfnamefont {E.~V.}\ \bibnamefont
  {Albano}},\ }\href@noop {} {\bibfield  {journal} {\bibinfo  {journal} {J.
  Phys. A}\ }\textbf {\bibinfo {volume} {27}},\ \bibinfo {pages} {L881}
  (\bibinfo {year} {1994})}\BibitemShut {NoStop}%
\bibitem [{\citenamefont {Grassberger}(1983)}]{Grassberger}%
  \BibitemOpen
  \bibfield  {author} {\bibinfo {author} {\bibfnamefont {P.}~\bibnamefont
  {Grassberger}},\ }\href@noop {} {\bibfield  {journal} {\bibinfo  {journal}
  {Math. Biosci.}\ }\textbf {\bibinfo {volume} {63}},\ \bibinfo {pages} {157}
  (\bibinfo {year} {1983})}\BibitemShut {NoStop}%
\bibitem [{\citenamefont {Zhou}\ \emph {et~al.}(2015)\citenamefont {Zhou},
  \citenamefont {Mowrey}, \citenamefont {Tang},\ and\ \citenamefont
  {Xu}}]{Zhou}%
  \BibitemOpen
  \bibfield  {author} {\bibinfo {author} {\bibfnamefont {D.~W.}\ \bibnamefont
  {Zhou}}, \bibinfo {author} {\bibfnamefont {D.~D.}\ \bibnamefont {Mowrey}},
  \bibinfo {author} {\bibfnamefont {P.}~\bibnamefont {Tang}}, \ and\ \bibinfo
  {author} {\bibfnamefont {Y.}~\bibnamefont {Xu}},\ }\href@noop {} {\bibfield
  {journal} {\bibinfo  {journal} {Phys. Rev. Lett.}\ }\textbf {\bibinfo
  {volume} {115}},\ \bibinfo {pages} {108103} (\bibinfo {year}
  {2015})}\BibitemShut {NoStop}%
\bibitem [{\citenamefont {Paul}(2009)}]{Paul}%
  \BibitemOpen
  \bibfield  {author} {\bibinfo {author} {\bibfnamefont {R.}~\bibnamefont
  {Paul}},\ }\href@noop {} {\bibfield  {journal} {\bibinfo  {journal} {Eur.
  Phys. J. E}\ }\textbf {\bibinfo {volume} {30}},\ \bibinfo {pages} {101}
  (\bibinfo {year} {2009})}\BibitemShut {NoStop}%
\bibitem [{\citenamefont {Araujo}\ \emph {et~al.}(2014)\citenamefont {Araujo},
  \citenamefont {Grassberger}, \citenamefont {Kahng}, \citenamefont {Schrenk},\
  and\ \citenamefont {Ziff}}]{Araujo}%
  \BibitemOpen
  \bibfield  {author} {\bibinfo {author} {\bibfnamefont {N.}~\bibnamefont
  {Araujo}}, \bibinfo {author} {\bibfnamefont {P.}~\bibnamefont {Grassberger}},
  \bibinfo {author} {\bibfnamefont {B.}~\bibnamefont {Kahng}}, \bibinfo
  {author} {\bibfnamefont {K.~J.}\ \bibnamefont {Schrenk}}, \ and\ \bibinfo
  {author} {\bibfnamefont {R.~M.}\ \bibnamefont {Ziff}},\ }\href@noop {}
  {\bibfield  {journal} {\bibinfo  {journal} {Eur. Phys. J. Spec. Top.}\
  }\textbf {\bibinfo {volume} {223}},\ \bibinfo {pages} {2307} (\bibinfo {year}
  {2014})}\BibitemShut {NoStop}%
\bibitem [{\citenamefont {Hoshen}\ \emph {et~al.}(1979)\citenamefont {Hoshen},
  \citenamefont {Klymko},\ and\ \citenamefont {Kopelman}}]{Hoshen}%
  \BibitemOpen
  \bibfield  {author} {\bibinfo {author} {\bibfnamefont {J.}~\bibnamefont
  {Hoshen}}, \bibinfo {author} {\bibfnamefont {P.}~\bibnamefont {Klymko}}, \
  and\ \bibinfo {author} {\bibfnamefont {R.}~\bibnamefont {Kopelman}},\
  }\href@noop {} {\bibfield  {journal} {\bibinfo  {journal} {J. Stat. Phys.}\
  }\textbf {\bibinfo {volume} {21}},\ \bibinfo {pages} {583} (\bibinfo {year}
  {1979})}\BibitemShut {NoStop}%
\bibitem [{\citenamefont {Achlioptas}\ \emph {et~al.}(2009)\citenamefont
  {Achlioptas}, \citenamefont {D’Souza},\ and\ \citenamefont
  {Spencer}}]{Achlioptas}%
  \BibitemOpen
  \bibfield  {author} {\bibinfo {author} {\bibfnamefont {D.}~\bibnamefont
  {Achlioptas}}, \bibinfo {author} {\bibfnamefont {R.~M.}\ \bibnamefont
  {D’Souza}}, \ and\ \bibinfo {author} {\bibfnamefont {J.}~\bibnamefont
  {Spencer}},\ }\href@noop {} {\bibfield  {journal} {\bibinfo  {journal}
  {Science}\ }\textbf {\bibinfo {volume} {323}},\ \bibinfo {pages} {1453}
  (\bibinfo {year} {2009})}\BibitemShut {NoStop}%
\bibitem [{\citenamefont {Ziff}(2009)}]{Ziff_explo}%
  \BibitemOpen
  \bibfield  {author} {\bibinfo {author} {\bibfnamefont {R.~M.}\ \bibnamefont
  {Ziff}},\ }\href@noop {} {\bibfield  {journal} {\bibinfo  {journal} {Phys.
  Rev. Lett.}\ }\textbf {\bibinfo {volume} {103}},\ \bibinfo {pages} {045701}
  (\bibinfo {year} {2009})}\BibitemShut {NoStop}%
\bibitem [{\citenamefont {Cho}\ \emph {et~al.}(2009)\citenamefont {Cho},
  \citenamefont {Kim}, \citenamefont {Park}, \citenamefont {Kahng},\ and\
  \citenamefont {Kim}}]{Cho}%
  \BibitemOpen
  \bibfield  {author} {\bibinfo {author} {\bibfnamefont {Y.~S.}\ \bibnamefont
  {Cho}}, \bibinfo {author} {\bibfnamefont {J.~S.}\ \bibnamefont {Kim}},
  \bibinfo {author} {\bibfnamefont {J.}~\bibnamefont {Park}}, \bibinfo {author}
  {\bibfnamefont {B.}~\bibnamefont {Kahng}}, \ and\ \bibinfo {author}
  {\bibfnamefont {D.}~\bibnamefont {Kim}},\ }\href@noop {} {\bibfield
  {journal} {\bibinfo  {journal} {Phys. Rev. Lett.}\ }\textbf {\bibinfo
  {volume} {103}},\ \bibinfo {pages} {135702} (\bibinfo {year}
  {2009})}\BibitemShut {NoStop}%
\bibitem [{\citenamefont {Adler}(1991)}]{Adler}%
  \BibitemOpen
  \bibfield  {author} {\bibinfo {author} {\bibfnamefont {J.}~\bibnamefont
  {Adler}},\ }\href@noop {} {\bibfield  {journal} {\bibinfo  {journal} {Physica
  A}\ }\textbf {\bibinfo {volume} {171}},\ \bibinfo {pages} {453} (\bibinfo
  {year} {1991})}\BibitemShut {NoStop}%
\bibitem [{\citenamefont {Makse}\ \emph {et~al.}(1995)\citenamefont {Makse},
  \citenamefont {Havlin},\ and\ \citenamefont {Stanley}}]{Makse}%
  \BibitemOpen
  \bibfield  {author} {\bibinfo {author} {\bibfnamefont {H.~A.}\ \bibnamefont
  {Makse}}, \bibinfo {author} {\bibfnamefont {S.}~\bibnamefont {Havlin}}, \
  and\ \bibinfo {author} {\bibfnamefont {H.~E.}\ \bibnamefont {Stanley}},\
  }\href@noop {} {\bibfield  {journal} {\bibinfo  {journal} {Nature}\ }\textbf
  {\bibinfo {volume} {377}},\ \bibinfo {pages} {608} (\bibinfo {year}
  {1995})}\BibitemShut {NoStop}%
\bibitem [{\citenamefont {Araujo}\ \emph {et~al.}(2002)\citenamefont {Araujo},
  \citenamefont {Moreira}, \citenamefont {Makse}, \citenamefont {Stanley},\
  and\ \citenamefont {Jr.}}]{Araujo2}%
  \BibitemOpen
  \bibfield  {author} {\bibinfo {author} {\bibfnamefont {A.~D.}\ \bibnamefont
  {Araujo}}, \bibinfo {author} {\bibfnamefont {A.~A.}\ \bibnamefont {Moreira}},
  \bibinfo {author} {\bibfnamefont {H.~A.}\ \bibnamefont {Makse}}, \bibinfo
  {author} {\bibfnamefont {H.~E.}\ \bibnamefont {Stanley}}, \ and\ \bibinfo
  {author} {\bibfnamefont {J.~S.}\ \bibnamefont {Andrade}},\ }\href@noop {}
  {\bibfield  {journal} {\bibinfo  {journal} {Phys. Rev. E}\ }\textbf {\bibinfo
  {volume} {66}},\ \bibinfo {pages} {046304} (\bibinfo {year}
  {2002})}\BibitemShut {NoStop}%
\bibitem [{\citenamefont {Kundu}\ and\ \citenamefont {Manna}(2016)}]{Kundu}%
  \BibitemOpen
  \bibfield  {author} {\bibinfo {author} {\bibfnamefont {S.}~\bibnamefont
  {Kundu}}\ and\ \bibinfo {author} {\bibfnamefont {S.~S.}\ \bibnamefont
  {Manna}},\ }\href@noop {} {\bibfield  {journal} {\bibinfo  {journal} {Phys.
  Rev. E}\ }\textbf {\bibinfo {volume} {93}},\ \bibinfo {pages} {062133}
  (\bibinfo {year} {2016})}\BibitemShut {NoStop}%
\bibitem [{\citenamefont {Hassan}\ and\ \citenamefont {Rahman}(2015)}]{Hassan}%
  \BibitemOpen
  \bibfield  {author} {\bibinfo {author} {\bibfnamefont {M.~K.}\ \bibnamefont
  {Hassan}}\ and\ \bibinfo {author} {\bibfnamefont {M.~M.}\ \bibnamefont
  {Rahman}},\ }\href@noop {} {\bibfield  {journal} {\bibinfo  {journal} {Phys.
  Rev. E}\ }\textbf {\bibinfo {volume} {92}},\ \bibinfo {pages} {040101(R)}
  (\bibinfo {year} {2015})}\BibitemShut {NoStop}%
\bibitem [{\citenamefont {Hoshen}\ and\ \citenamefont {Kopelman}(1976)}]{HK}%
  \BibitemOpen
  \bibfield  {author} {\bibinfo {author} {\bibfnamefont {J.}~\bibnamefont
  {Hoshen}}\ and\ \bibinfo {author} {\bibfnamefont {R.}~\bibnamefont
  {Kopelman}},\ }\href@noop {} {\bibfield  {journal} {\bibinfo  {journal}
  {Phys. Rev. B}\ }\textbf {\bibinfo {volume} {14}},\ \bibinfo {pages} {3438}
  (\bibinfo {year} {1976})}\BibitemShut {NoStop}%
\end{thebibliography}
\end{document}